\documentclass[final,5p,times,twocolumn,compress]{elsarticle}
\usepackage{amssymb}
\usepackage{amsmath,amsfonts}
\usepackage{algorithmic}
\usepackage{algorithm}
\usepackage{array}
\usepackage[caption=false,font=normalsize,labelfont=sf,textfont=sf]{subfig}
\usepackage{textcomp}
\usepackage{stfloats}
\usepackage{url}
\usepackage{verbatim}
\usepackage{graphicx}
\usepackage{cite}
\usepackage{textcomp}
\usepackage {upgreek}
\usepackage{subfloat}
\usepackage[percent]{overpic}
\usepackage{graphics}
\usepackage{caption}
\usepackage{adjustbox}
\usepackage{hyperref}
\usepackage{xcolor}
\usepackage{booktabs} 

\journal{Optics \& Laser Technology}

\begin{document}
\captionsetup[figure]{labelfont={bf},labelformat={default},labelsep=period,name={Fig.}}
\begin{frontmatter}

\title{Frequency-shifted laser feedback interferometry in non-planar ring oscillators}

\author[1,2]{Rong Zhu\fnref{equal}} 
\author[1]{Xuezhen Gong\fnref{equal}}
\author[1]{Wenxun Li}
\author[1]{Guobin Zhou}
\author[1]{Weitong Fan}
\author[1]{Danqing Liu}
\author[1]{Chunzhao Ma}
\author[1]{Jie Xu}
\author[1]{Changlei Guo\corref{cor1}}
\ead{guochlei@mail.sysu.edu.cn}
\author[1]{Hsien-Chi Yeh}





\affiliation[1]{organization={MOE Key Laboratory of TianQin Mission, TianQin Research Center for Gravitational Physics \& School of Physics and Astronomy, Frontiers Science Center for TianQin, CNSA Research Center for Gravitational Waves}, 
                addressline={Sun Yat-sen University (Zhuhai Campus)}, 
                postcode={519082}, 
                city={Zhuhai},
                country={China}}

\affiliation[2]{organization={Department of Physics, City University of Hong Kong}, 
                addressline={Tat Chee Avenue}, 
                , 
                postcodesep={}, 
                city={Hong Kong SAR},
                country={China}}

\cortext[cor1]{Corresponding author.}
\fntext[equal]{The two authors contribute equally to this work.}

\begin{abstract}
Laser feedback interferometry (LFI) has a wide range of applications such as displacement, distance and velocity measurements. LFI has been realized in many types of lasers but has never been reported in non-planar ring oscillators (NPRO) to the best of our knowledge. Here, we present a new type of LFI based on an NPRO laser. The intrinsic resistance to optical feedback in NPROs is broken under weak-magnetic-intensity condition, where stable bidirectional lasing is initiated in the ring cavity. The interference signal, i.e., the beat of the bidirectional lasing is with frequency in the range of a few hundred kilohertz, which is mainly determined by the applied magnetic intensity in NPRO. Frequency-shifted LFI is thus constructed in NPRO without using acoustic-optic modulators as mostly used in conventional LFI. A theoretical model based on two-frequency rate equations and Lang–Kobayashi equation is presented to describe the mechanism of LFI in NPRO. In the end, micro-vibrational measurements are demonstrated to prove the potential application, where vibration-detection amplitude limit is sub-picometer, and the detection frequency range from kilohertz to a few hundred kilohertz is achieved. Benefiting from the characteristics of tiny footprint, ruggedized structure, long lifetime and ultralow-noise of NPRO lasers, NPRO-based LFI may find important applications in industry, scientific research, military and aerospace.
\end{abstract}

\begin{keyword}
Laser feedback interferometry, non-planar ring oscillator, bidirectional lasing.
\end{keyword}

\end{frontmatter}



\section{Introduction}

Laser feedback interferometry (LFI) has been widely used in metrology \citep{Tian:24,2}, laser parameters measurement \citep{2,3,4}, physical quantities measurement and other fields \citep{5,6} for its simple structure, easy alignment and high sensitivity \citep{7,8,9}. Plenty of theoretical models about LFI have also been studied, varying from single-mode lasers like solid-state \citep{14} and  semiconductor lasers \citep{27}, to multi-mode lasers such as quantum cascade lasers  \citep{28,29}. The measuring principle is based on self-mixing effect which uses re-injected photons back into the laser cavity to produce informative modulations of the laser intensity, frequency, phase or even polarization. Frequency-shifted LFI has been demonstrated to show better performance than heterodyne interferometry for a wide range of laser powers and detection noise levels \citep{10,11,12}. This performance can be attributed to the amplification gain inside the laser cavity of the LFI, which is on the order of 10$^{3}$ to 10$^{6}$ in Class-B lasers \citep{14,13}. In the meantime, the shifted frequency (close to, in resonance with or much higher than the laser relaxation-oscillation (RO) frequency) gives a freedom to be tuned to find the best measurement resolution or sensitivity \citep{11}. Accordingly, picometer displacement resolutions \citep{7}, sub-picowatt \citep{8} or even single-photon sensitivities \citep{9}, have been demonstrated in solid-state microchip laser and fiber laser systems. However, because the round-trip resonant frequencies are degenerate in most reciprocal laser resonators (i.e. Fabry-Perot or ring resonators without nonreciprocal elements), the required frequency shifts (from kilohertz to megahertz) have to rely on external devices, i.e. a pair of acoustic-optic modulators (AOMs) \citep{14,12,13,15,16,17}. This evidently adds complexity and cost to the system.

Non-planar ring oscillators (NPROs) \citep{18}, with applied magnetic field, have non-degenerate resonant frequencies through nonreciprocal rotations of the polarizations in their clockwise and counterclockwise roundtrips \citep{19}. The resonant frequency differences from kilohertz to megahertz can be exploited as intrinsic interference signal for LFI in NPRO laser. However, this phenomenon has never been reported before, to the best of our knowledge. The reason might be that the NPROs are in most cases working under high-intensity magnetic field (corresponding to high loss-differences), which makes them resistant to optical feedback \citep{20}. Recently, we found that the loss-difference required for unidirectional lasing is not as large as a former empirical value of 0.01\%, but could be as low as 0.0001\% \citep{21}. The lowered magnetic intensity decreases its resistance to optical feedback and leads to the observation of frequency-shifted LFI in NPROs. Since the interference signal is engendered by the beat signal of two modes inside NPRO, without using external frequency-shifted devices such as AOMs, the system is quite simple compared with traditional frequency-shifted LFI. 

In this work, we demonstrate a new type of frequency-shifted LFI in NPROs. Dual-frequency rate equations \citep{22} with the Lang–Kobayashi equation \citep{23} are introduced to describe the feedback mechanism in NPRO. In the model, the same modulation term in two stationary solutions proves the theoretical validity of LFI in NPRO. In the end, LFI experiment is set up to demonstrate the application. The interference signal which comes from the beat of bidirectional lasing is measured and discussed. Moreover, vibration amplitudes from tens of picometers to a few nanometers with vibration frequency from kilohertz to a few hundred kilohertz are detected, and the detection limit (noise floor) is sub-picometer in the NPRO LFI system. We believe our theory and experiment break a new path for frequency shifted LFI in NPROs. 

\section{Mechanism and principle}

\begin{figure}[h]
\centering
\includegraphics[width=3.5in]{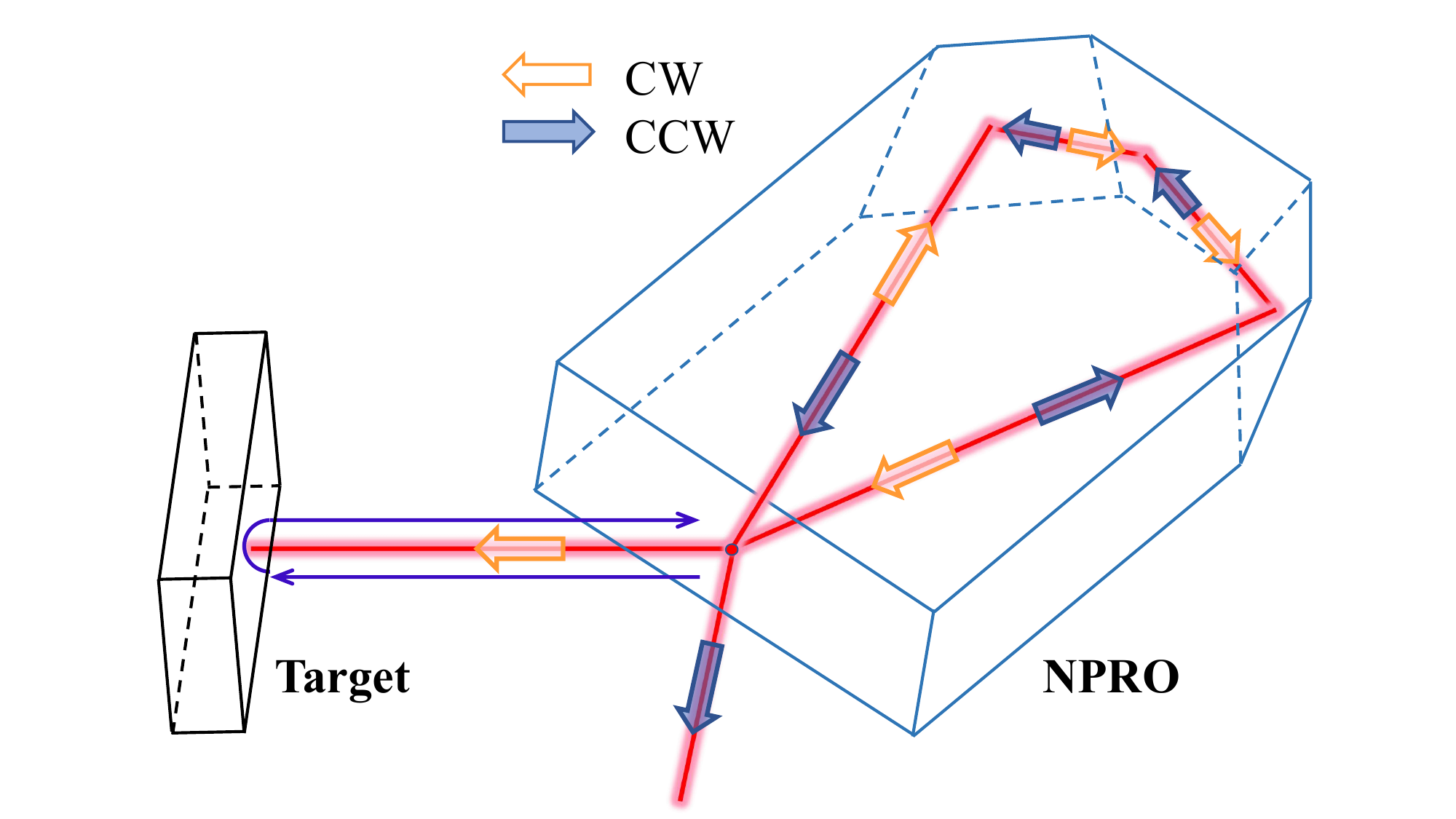}
\caption{The schematic of monolithic NPRO with optical feedback. The CW and CCW modes are shown when the feedback target is at CW path.}
\label{Fig1}
\end{figure}

Under weak magnetic intensity conditions, the loss difference between the first two low-loss eigenmodes in NPRO is small that the resistance to optical feedback is weaken \citep{20}. The re-injected light from the CW/CCW (CW, clockwise, CCW, counter-clockwise) mode is coupled to the other mode, as shown schematically in Fig.\ref{Fig1}. As a result, the bidirectional lasing in both CW and CCW directions is initiated in NPRO. Due to the mode coupling, the CW and CCW laser information is observed in each of the CW or CCW direction. This describes the basic physical phenomenon of LFI in NPRO. To theoretically explain the phenomenon, we consider the NPRO laser system (Nd:YAG with 808 nm laser pumping and 1064 nm signal laser in our case) as a two-level system. Then we define $N_{CW}$ and $N_{CCW}$ as the population inversion of CW and CCW modes, respectively. Combined with Lang–Kobayashi equation \citep{23}, the complex forms of the two-frequency laser equations are given by:
\begin{equation}
\label{eq1}
\begin{aligned}
\frac{d}{dt} E_{CW}(t) e^{i\omega_1 t} = \left[i\omega_{1} + \frac{1}{2} (\varepsilon N_{CW} - \gamma_{CW})\right] E_{CW}(t) e^{i\omega_1 t}
\end{aligned}
\end{equation}

\begin{equation}
\label{eq2}
\begin{aligned}
\frac{d}{dt} E_{CCW}(t) e^{i\omega_2 t} = \left[i\omega_{2} + \frac{1}{2} (\varepsilon N_{CCW} - \gamma_{CCW})\right] E_{CCW}(t) e^{i\omega_2 t} \\
+ \gamma_{\text{ext}} E_{CW}(t-\tau) e^{i\omega_1 (t-\tau)}
\end{aligned}
\end{equation}

\begin{equation}
\label{eq3}
\begin{aligned}
\frac{dN_{CW}}{dt} = \gamma [N_{01} - N_{CW}] - \varepsilon N_{CW} \left[ |E_{CW}(t)|^2 + \xi_{12} |E_{CCW}(t)|^2 \right]
\end{aligned}
\end{equation}

\begin{equation}
\label{eq4}
\begin{aligned}
\frac{dN_{CCW}}{dt} = \gamma [N_{02} - N_{CCW}] - \varepsilon N_{CCW} \left[ |E_{CCW}(t)|^2 + \xi_{21} |E_{CW}(t)|^2 \right]
\end{aligned}
\end{equation}

\noindent
Here, we assume that the re-injected light is CW mode. The laser cavity frequency of CW and CCW modes are $\omega_{1}$ and $\omega_{2}$, respectively. $E_{CW}$ and $E_{CCW}$ represent the complex amplitude of the two electric fields, $\gamma$ is the population inversion decay rate. $\gamma_{CW}$ and $\gamma_{CCW}$ are the decay rates of laser cavity in CW and CCW modes, which are assumed to be different because of the non-reciprocity of NPRO \citep{19}. $\epsilon$ is the excited emission-coefficient which is related to the Einstein coefficient B. To describe the coupling between CW and CCW modes, coefficients $\xi_{12}$ and $\xi_{21}$ are introduced as the ratio of the cross- to the self-saturation coefficient, and $\beta = \xi_{12} \xi_{21}$ represents the nonlinear coupling constant \citep{22}. $N_{01}$ and $N_{02}$ are the unsaturated population inversion of CW and CCW modes, respectively. $\gamma_{\text{ext}}$ is re-injection rate of feedback electric field which is related to the reflectivity of the target and coupling inside the cavity \citep{14}, and $\tau = \frac{2L}{c}$ is the photon time delay between the laser and target, and $L$ is the distance between them (for simplicity, the refractive index of the medium is 1).

First, we consider the case when the system has no feedback ($\gamma_{\text{ext}}=0$ in Eq.\eqref{eq2}). We then have the stationary solutions of population inversions $N_{s1}$, $N_{s2}$, as well as the stationary intensities $I_{s1}$ and $I_{s2}$, corresponding to CW and CCW modes, respectively:

\begin{equation}
\label{eq5}
\begin{aligned}
 N_{s1} = \frac{\gamma_{CW}}{\varepsilon}
\end{aligned}
\end{equation}

\begin{equation}
\label{eq6}
\begin{aligned}
 N_{s2} = \frac{\gamma_{CCW}}{\varepsilon}
\end{aligned}
\end{equation}

\begin{equation}
\label{eq7}
\begin{aligned}
 I_{s1} = \left| E_{CW}(t) \right|^2 = \frac{\gamma (1 - \xi_{12})}{\varepsilon (1 - \beta)} \left( \frac{\varepsilon N_0}{\gamma_{CW}+\gamma_{CCW}} - 1 \right)
\end{aligned}
\end{equation}

\begin{equation}
\label{eq8}
\begin{aligned}
 I_{s2} = \left| E_{CCW}(t) \right|^2 = \frac{\gamma (1 - \xi_{21})}{\varepsilon (1 - \beta)} \left( \frac{\varepsilon N_0}{\gamma_{CW}+\gamma_{CCW}} - 1 \right)
\end{aligned}
\end{equation}

In NPRO system, CW and CCW laser modes are from the same population inversion excitation. The sum of the lasing power in two directions always be the same because of the conservation principle. We thus define the total unsaturated population inversion $N_0$ of NPRO as the sum of two directions, $N_0 = N_{01} + N_{02}$. It is seen as a constant under the same pumping condition and does not vary as a function of time or magnetic intensity. And we define $r_1 = \frac{N_{01}}{N_{s1}}$ and $r_2 = \frac{N_{02}}{N_{s2}}$ which are the corresponding to excitation ratios for two modes \citep{24}. Since the bidirectional laser modes share the same population inversion excitation, we consider the two excitation ratios to be equal, which means $r_1=r_2=r$. Then we get the relation between two unsaturated population inversions:

\begin{equation}
\label{eq9}
\begin{aligned}
 N_{01} = \frac{\gamma_{CW}}{\gamma_{CW} + \gamma_{CCW}} N_0
\end{aligned}
\end{equation}

\begin{equation}
\label{eq10}
\begin{aligned}
 N_{02} = \frac{\gamma_{CCW}}{\gamma_{CW} + \gamma_{CCW}} N_0
\end{aligned}
\end{equation}
 
With optical feedback, the stationary intensity solutions $I_{s1}'$ and $I_{s2}'$ are obtained from Eqs.\eqref{eq1}-\eqref{eq4}. For the interests of simplicity, we consider the case of the slowly varying amplitude and the photon time delay $\tau $ is much shorter than the period of the beat signal between the two modes, which implies $E(t - \tau) \approx E(t)$.  Then we have:

\begin{equation}
\label{eq11}
\begin{aligned}
I_{s1}' = \frac{\gamma (1 - \xi_{12})}{\varepsilon (1 - \beta)} \left( \frac{\varepsilon N_0}{\gamma_{CW} + \gamma_{CCW} - 2 \gamma_{ext} \cos(\Delta \omega t - \omega_1 \tau)} - 1 \right)\\
\approx I_{s1} + \frac{rI_{s1}}{r-1} \frac{\gamma_{ext}}{\gamma_B} \cos(\Delta \omega t - \omega_1 \tau) 
\end{aligned}
\end{equation}

\begin{equation}
\label{eq12}
\begin{aligned}
I_{s2}' = \frac{\gamma (1 - \xi_{21})}{\varepsilon (1 - \beta)} \left( \frac{\varepsilon N_0}{\gamma_{CW} + \gamma_{CCW} - 2 \gamma_{ext} \cos(\Delta \omega t - \omega_1 \tau)} - 1 \right)\\
\approx I_{s2} +  \frac{rI_{s2}}{r-1} \frac{\gamma_{ext}}{\gamma_B} \cos(\Delta \omega t - \omega_1 \tau) 
\end{aligned}
\end{equation}
where $\Delta \omega = \omega_2 - \omega_1$ and $\gamma_{B}=\gamma_{CW}+\gamma_{CCW}$, and we consider the case when $\gamma_{ext} \tau \ll 1$. From Eqs.\eqref{eq11}-\eqref{eq12} we can see that after introducing the external feedback term to the laser equations, the intensity solutions become the steady solutions with no feedback Eqs.\eqref{eq7}-\eqref{eq8} plus a modulation term with target feedback signal, and this modulation term shows up in both solutions. The periodical amplitude modulation is equivalent to a beat signal at angular frequency $\Delta \omega$, and its phase is directly related to the time delay $\tau$ of re-injected photons. When the distance between the laser and the target changes by  $\Delta L$, the change in the phase of beat signal is:
\begin{equation}
\label{eq13}
\begin{aligned}
\Delta \varphi  = {\omega _1}\Delta \tau  = \frac{{2\pi c}}{\lambda } \cdot \frac{{2\Delta L}}{c}=\frac{{4\pi }}{\lambda }\Delta L
\end{aligned}
\end{equation}
with $\lambda$ the laser wavelength. This equation proves that the NPRO feedback system can achieve self-feedback interferometry with an interference signal acting as the amplitude modulation in both sides. 

\section{Experiment and results}

\begin{figure}[h]
\centering
\includegraphics[width=3.5in]{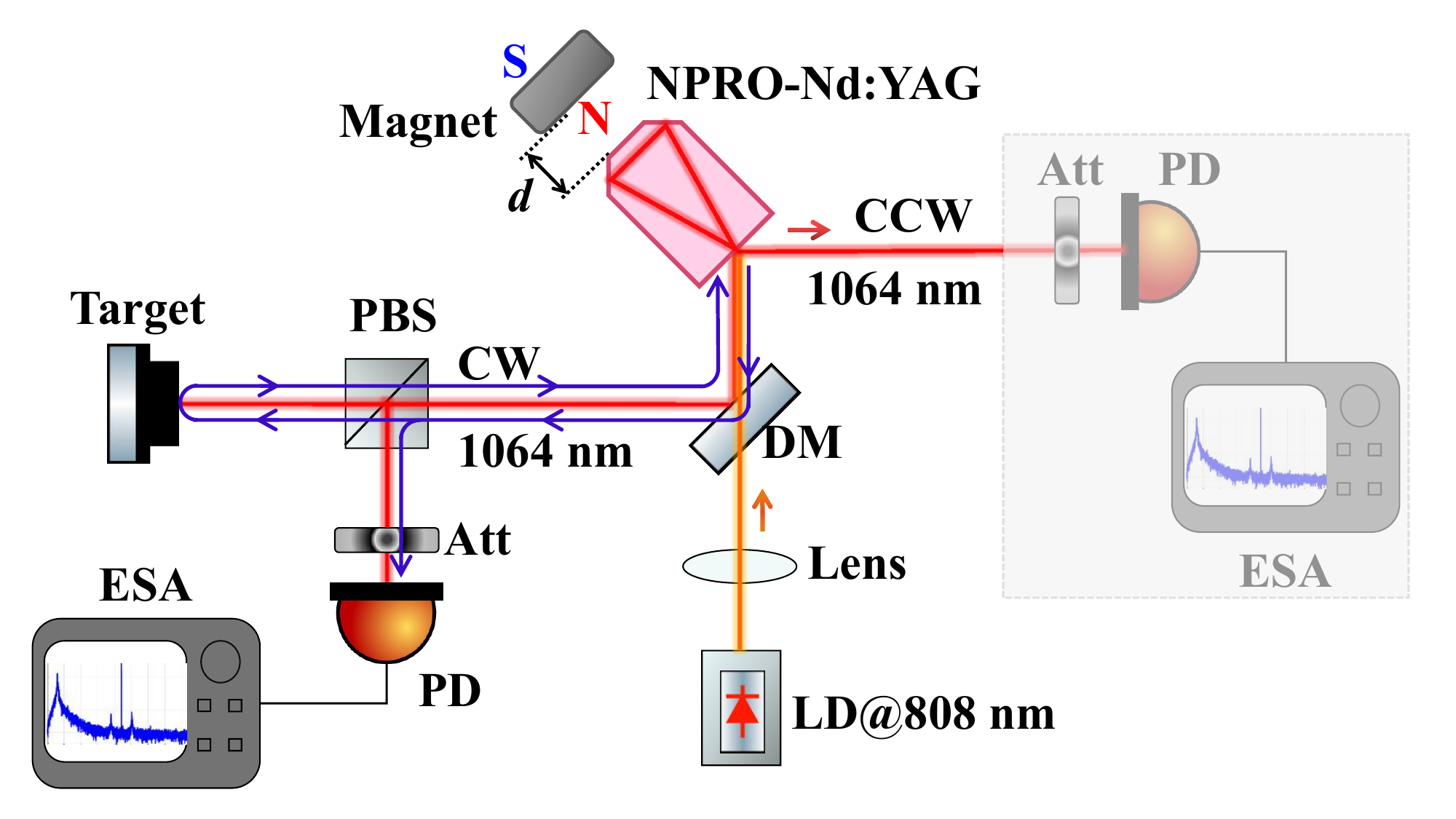}
\caption{Schematic of the experimental setup for the NPRO LFI system. \textit{d} is the distance between the magnet and the NPRO. N and S represent the magnetic poles. LD, laser diode; DM, dichroic mirror with high transmittance at 808 nm and high reflection at 1064 nm; CW, clockwise; CCW, counter-clockwise; PBS, polarization beam splitter; Att, optical attenuator; PD, photodetector; ESA, electronic spectrum analyzer. The shadow zone on the CCW light path can realize similar measurements as that on the CW and is not used in this study. Orange arrow, 808 nm pumping laser. Red arrow, 1064 nm CCW mode. Blue arrows, 1064 nm CW mode  when introduced optical feedback.}
\label{Fig2}
\end{figure}

To verify the theoretical predictions of the LFI in NPRO, we build an experiment as shown schematically in Fig.\ref{Fig2}. Laser diode operating at 808 nm wavelength is the pumping source for NPRO, where Nd:YAG crystal is the gain material. A lens (f = 15 mm) is used to focus the pump light onto the NPRO. The NPRO has a dimension of 3 mm × 8 mm ×12 mm and exhibits an incident angle of 30°, an out-of-plane angle of 90°. The front surface of the crystal is coated with high-transmittance at 808 nm and high-reflectance (99.8\% for S polarization and 94\% for P polarization) at 1064 nm. The NPRO operates under temperature around 25℃ with an active temperature controller. The distance \textit{d} between the permanent magnet and the NPRO is controlled by a one-dimensional transition-stage. In this way, the magnetic intensity applied on the NPRO can be tuned with \textit{d}. The direction of magnetic field is chosen for CW emission under unidirectional condition. With the distance \textit{d} increase, the NPRO shows weak resistance to feedback and generates a beat signal between the CW and CCW modes themselves. The LFI experiment is operated under this circumstance when the feedback light oscillates inside the cavity and the interference signal appears in the outputs of two directions, as demonstrated in Eqs.\eqref{eq11}-\eqref{eq12}. More details on the experimental parameters are described in \citep{21} and \citep{24}. A dichroic mirror is placed between the laser diode and the NPRO, with high transmittance at 808 nm and high reflectance at 1064 nm. This mirror will reflect the CW laser towards an external target and simultaneously collect the feedback signal from the target to the NPRO. Between the target and the dichroic mirror, a polarization beam splitter is used to split part of the light for interference measurements, where the light is attenuated before detection. A low-noise photodetector with transimpedance amplifier (Thorlabs, PDA05CF2) is used here, and the electronic signal is processed by an electronic spectrum analyzer (Rohde \& Schwarz, FSV3007). On the CCW light path, a similar measurement (shadow zone in Fig.\ref{Fig2}) can also be constructed. We found the measurement sensitivity between the two paths is slightly different, which will be investigated in future study. In the next, the measurements are taken on the CW light path.

\begin{figure}[h]
\centering
\includegraphics[width=3in]{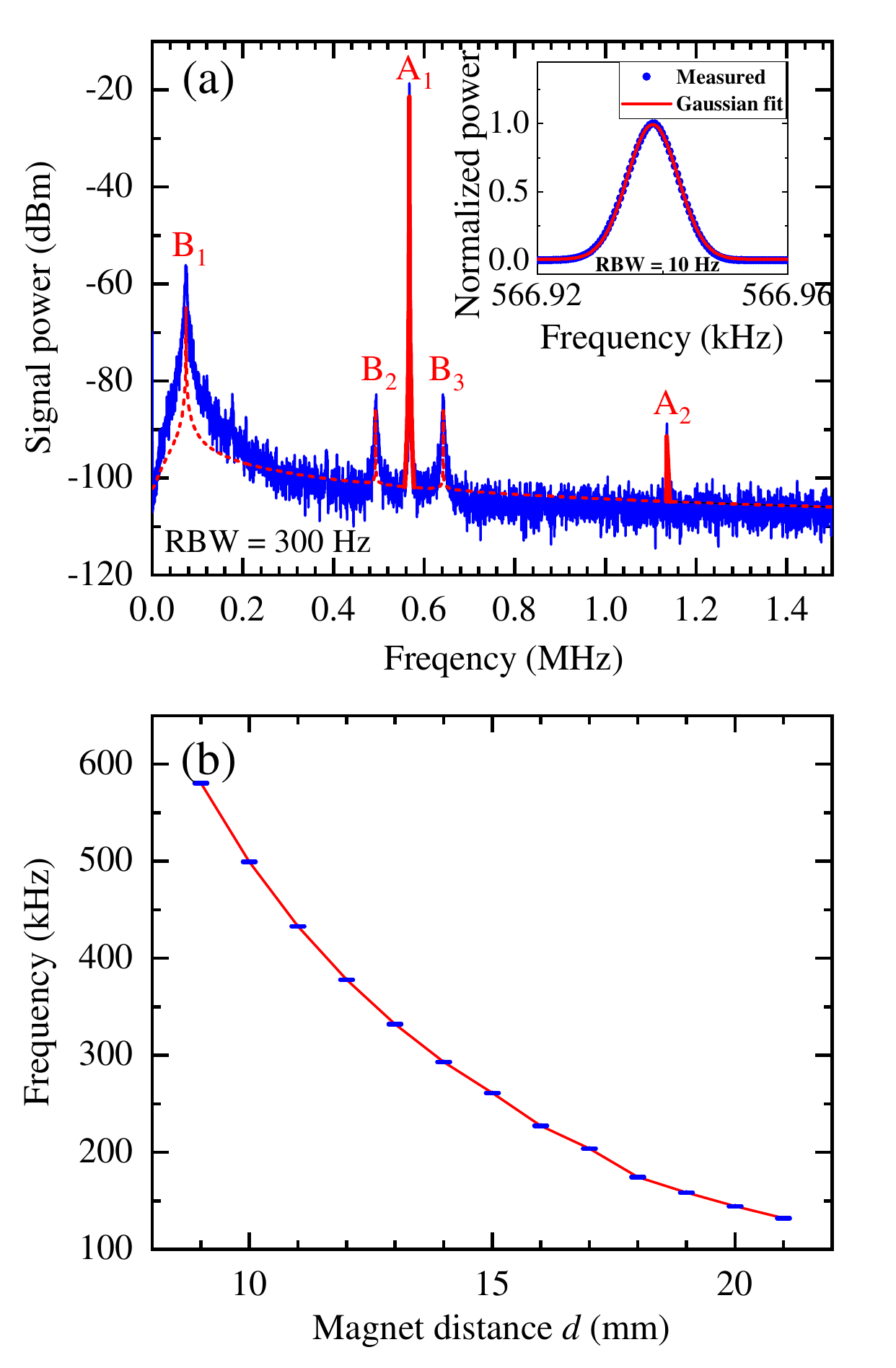}
\caption{ (a) Electronic spectrum measured as power of a typical beat signal (A\textsubscript{1}, centered at 567 kHz) with the NPRO RO noises (B\textsubscript{1}, centered at 74 kHz). A\textsubscript{2}: Secondary harmonic of A\textsubscript{1}. B\textsubscript{2} and B\textsubscript{3}: Sidebands of the beat signal A\textsubscript{1}, caused by RO B\textsubscript{1}. Blue curve: experiment data. Red solid curve: simulation of the NPRO feedback model. Red dashed curve: simulation of the RO model. The RBW is 300 Hz. Inset: Gaussian fitting of the normalized beat signal A\textsubscript{1} spectrum with RBW of 10 Hz. (b) Experimentally measured beat frequency versus magnet distance \textit{d}. The data is with error bars for three times of repeated measurement.}
\label{Fig3}
\end{figure}

A high-reflection (\textgreater99.5\%@1064 nm) mirror is firstly used as the target, whose position is finely adjusted to find the strongest beat signal. The mirror is decimeters far away from the NPRO and the CW and CCW light powers are as high as tens of milliwatt (under a pump power of 400 mW at 808 nm). A typical beat signal measured by the electronic spectrum analyzer is shown by the relative amplitude density in Fig.\ref{Fig3}(a), where the resolution bandwidth (RBW) is 300 Hz. The beat signal has a center frequency of 567 kHz. The two sidebands around it as well as the strong noise peak at around 74 kHz are from the RO of the NPRO laser itself. From Fig.\ref{Fig3}(a) one can see that the beat signal has a good signal-to-noise ratio. It is worth mentioning that investigation has also been conducted including two mirror-targets at the CW and CCW paths simultaneously. The result shows a chaos in the spectrum where the interference signal becomes unstable and sometimes even disappears. The reason might be that the mode competition between the CW and CCW modes becomes intense with two mirror-targets forming as an external cavity. Such investigation is still going on but is not the topic of this work.

\begin{figure}[h]
\centering
\includegraphics[width=3in]{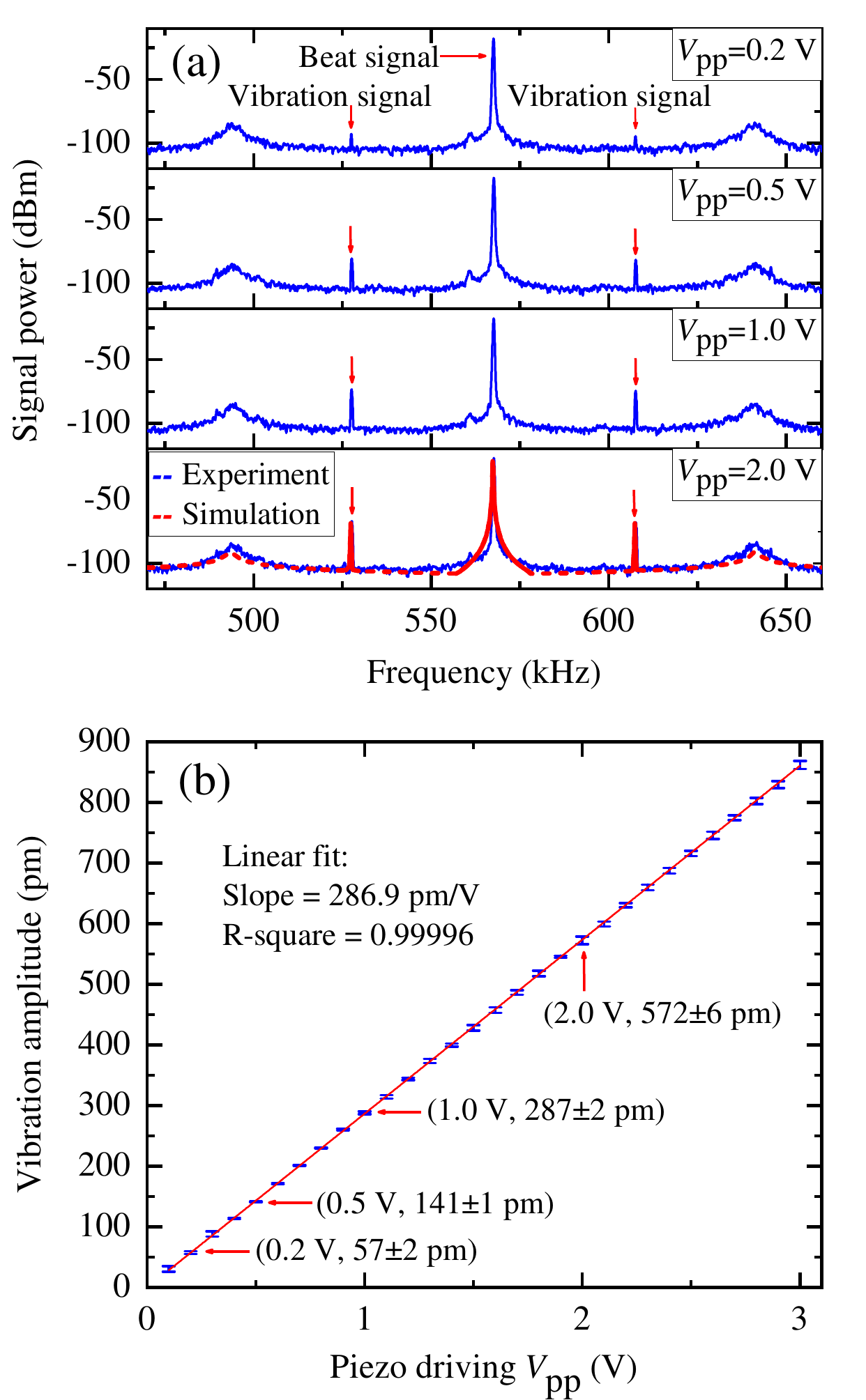}
\caption{ (a) Electronic spectra (measured as signal power) of the vibration detection with the beat signal (centered at 567 kHz), with the vibration signals (two thin sidebands) and the laser RO noises (two broad sidebands). The corresponding piezo driving \textit{V\textsubscript{pp}} of 0.2 V, 0.5 V, 1.0 V and 2.0 V are shown, respectively. Blue curve: experiment data. Red solid curve: simulation of the NPRO feedback model. Red dashed curve: simulation of the RO model. The RBW is 300 Hz. (b) Resolved vibration amplitude versus piezo driving voltage amplitude \textit{V\textsubscript{pp}}. The vibration frequency is at 40 kHz. Blue dots with error bars are from experimental measurements, red line is linear fit. Each point is averaged by five times of the left and right vibration signals. The goodness of the linear fit is 0.99945.}
\label{Fig4}
\end{figure}

We perform the simulation of NPRO feedback system by performing the Fourier transform for Eqs.\eqref{eq11}-\eqref{eq12}, and the results are shown in red solid lines of Fig.\ref{Fig3}(a) and Fig.\ref{Fig4}(a). Noticing that the intensity spectrum of NPRO has an unavoidable noise peak caused by RO. The model of RO has been studied well for decades\citep{30}. In this work, the characteristic of RO is not considered in the feedback model, but is contained in the stationary solutions of Eqs.\eqref{eq7}-\eqref{eq8} and the first term of Eqs.\eqref{eq11}-\eqref{eq12} for the experiment data. Thus, the RO model is combined in the simulation shown in red dashed lines of Fig.\ref{Fig3}(a) and Fig.\ref{Fig4}(a)) to provide the noise background of the spectrum. More details about the simulation can be found in the Appendix A.

To investigate the linewidth of the beat signal, we decrease the RBW from 500 to 10 Hz before the peak power starts to decrease. The inset in Fig.\ref{Fig3}(a) shows a normalized beat signal spectrum, where a Gaussian fit gives a linewidth of 10 Hz (the Gaussian curve is also the function of the window band-pass filter of the spectrum analyzer). Thus, it proves that the linewidth of the beat signal is no more than 10 Hz. Further frequency stability measurement using a frequency counter shows that the frequency noise of the beat signal is as low as 1 Hz/Hz$^{\frac{1}{2}}$ with Fourier frequency higher than 1 Hz (not shown here), which is a good indication for precision measurement applications.

Fig.\ref{Fig3}(b) shows the beat frequency versus magnet distance \textit{d}. By scanning \textit{d} from 9 mm to 21 mm one can see that the beat signal frequency is decreased from ~580 kHz to ~130 kHz. We have tested a few NPRO samples and found the beat frequencies are mostly stable in the range of hundreds of kilohertz. The lowest frequency can go to the low frequency side of the RO peak. Keeping increasing the distance will lead to competition between the two laser modes, and the beat signal becomes unstable \citep{21} or even disappears. With this magnetic-dependence, the SNR of interference signal shows an optimal area, but it varies between different NPRO samples, so we decide not to show here. The optimal magnet distance in our experiment is around 9.5 mm which gives a SNR more than 80 dB. This unique characteristic provides a method for the NPRO LFI system to operate under optimized conditions and improve the sensitivity of interference signal.

In conventional LFI, when the feedback power ratio is higher than a certain value (typically -30 dB or lower), chaos may emerge in the LFI system \citep{2}, which significantly deteriorates the applications. We found this situation does not happen in our NPRO LFI system. We estimated the feedback power ratio which is more than -10 dB (10\%) by measuring the optical power on both the CW and CCW paths, before and after the LFI is built. The obtained value (10\%) is far beyond the typical feedback levels where semiconductor-based LFI systems operate interferometrically. This indicates a significant difference between the NPRO LFI and some other LFI systems, especially semiconductor sources LFI. We further calculated the feedback level C \citep{25} by substituting parameters of our system (where the linewidth enhancement factor $\alpha$ is 1 for Nd:YAG crystal), which gives a value of ~0.1. The NPRO LFI characterization with different feedback power ratios and C values will be studied in the future. 

\begin{figure}[h]
\centering
\includegraphics[width=3in]{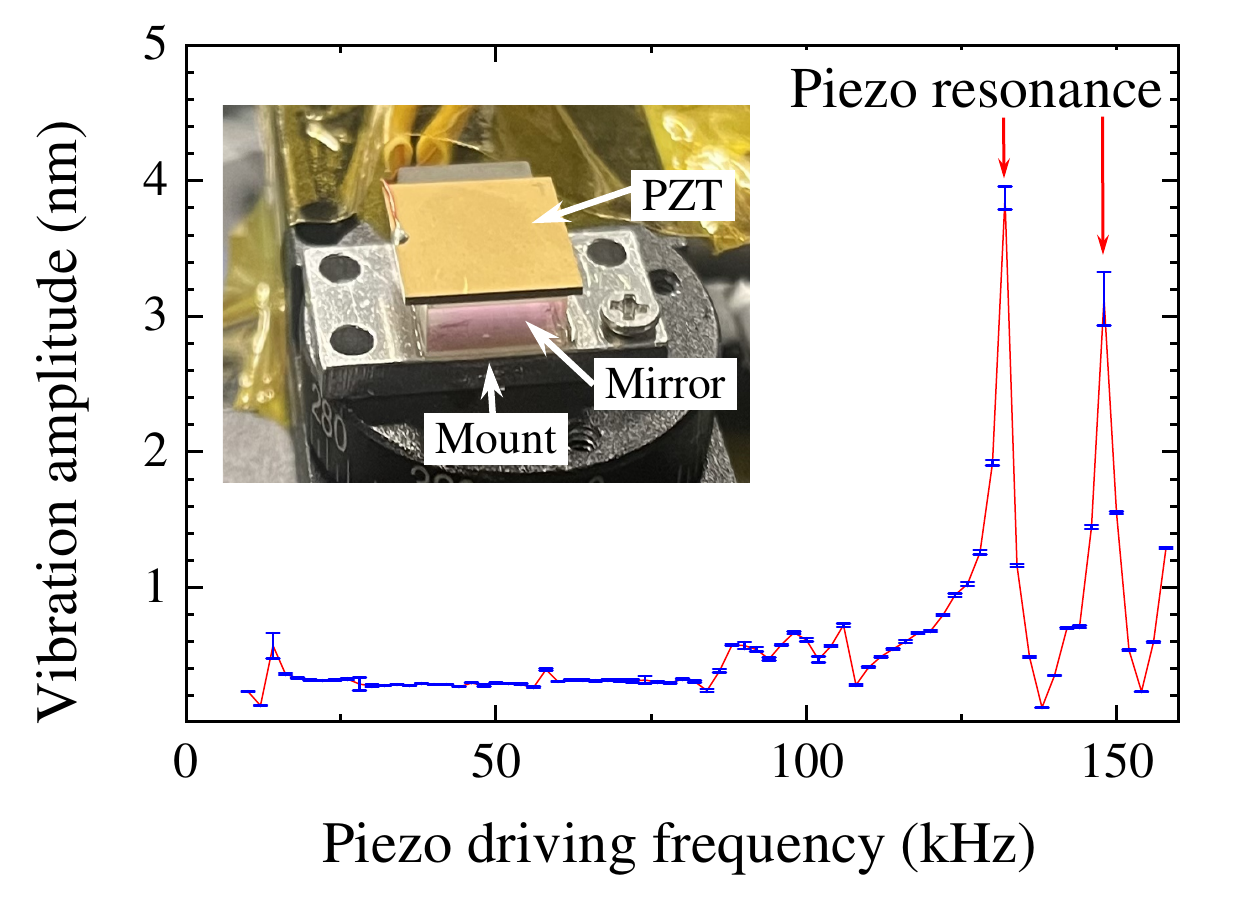}
\caption{ Vibration amplitude versus piezo driving frequency at constant piezo driving \textit{V\textsubscript{pp}}. The resonance peaks at around 132 kHz and 149 kHz are from the piezo itself. Inset: The piezo-mirror with mount. }
\label{Fig5}
\end{figure}

To demonstrate the application, we then replace the target with a piezo-mirror (as will be shown by the inset of Fig.\ref{Fig5}), where the mirror adhesive to a single-layer piezoelectric ceramic can generate a micro-vibration up to hundreds of kilohertz. The piezo is driven by a sine-wave voltage from a function generator. The oscillating mirror will simultaneously change the optical path length between the target and the NPRO.  As reflected in Eq.\eqref{eq13} , the vibration of the target introduces a periodic phase modulation in the beat signal:
\begin{equation}
\label{eq14}
\begin{aligned}
\Delta \varphi (t) = \frac{{4\pi}}{\lambda }A_{vib}\sin ({\omega _{vib}}t + \phi )
\end{aligned}
\end{equation}
Here, $A_{vib}$ amplitude, $\omega_{vib}$ frequency, $\phi$  original phase are the parameters of the target vibration respectively. Furthermore, this change is detected by the photodetector and processed by the electronic spectrum analyzer. When $A_{vib}$ is far smaller than $\lambda$, the first-order approximation of the Bessel functions contains three terms, corresponding with three peaks in the frequency spectrum: the beat signal in angular frequency $\Delta\omega$ with amplitude ${R_b}$, and two sidebands in angular frequency $\Delta\omega+\omega _{vib}$, $\Delta\omega-\omega _{vib}$, with the same amplitude ${R_s}$. The vibration amplitude $A _{vib}$ is then related to the ratio of sideband amplitude ${R_s}$ and beat signal amplitude ${R_b}$ \citep{26} as:
\begin{equation}
\label{eq15}
A_{vib} = \frac{\lambda}{2\pi} \frac{R_s}{R_b}
\end{equation}

Fig.\ref{Fig4}(a) shows typical electronic spectra when a 40 kHz sine-wave voltage modulation is applied on the piezo, where the corresponding piezo driving \textit{V\textsubscript{pp}} (peak to peak value) are respective 0.2 V, 0.5 V, 1.0 V and 2.0 V, and the RBW is 300 Hz. The two thin sidebands symmetrically distribute around the beat signal (at 567 kHz) with a 40 kHz frequency difference, which proves that the sine-wave modulation induced by micro-vibration has been detected by the NPRO LFI system. 

As the driving voltage amplitude is tuned from 0.1 V to 3.0 V, the corresponding electronic spectra are collected. 
 Finally, the vibration amplitude versus the applied voltage \textit{V\textsubscript{pp}} are resolved as shown in Fig.\ref{Fig4}(b), where each point is averaged five times of the left and right vibration signals.  Linear fitting gives a goodness of $R^2 = 0.99945$, which means that this detection works in a very good linear range. Note that these performances are comparable to that shown in \citep{8}, where AOMs are used to shift the beat frequency.

We then scanned the piezo driving frequency from 10 kHz to 160 kHz with constant \textit{V\textsubscript{pp}} (1.0 V) and calculated the corresponding vibration amplitudes. The typical
results are shown in Fig.\ref{Fig5}, where each point is averaged five times. One can see that the vibration amplitudes vary from 0.1 nm to 4 nm, and a few resonance (anti-resonance) peaks appear, which characterize the piezo-mirror. The lower bound of the vibration detection frequency is limited by the stability of the beat frequency, which can reach 1 kHz with the current setup. The upper bound of the vibration detection frequency is limited by the piezo itself used in the experiment. We have briefly increased the driving frequency to more than 500 kHz, and the vibration signals can still be seen.  

\begin{figure}[h]
\centering
\includegraphics[width=3in]{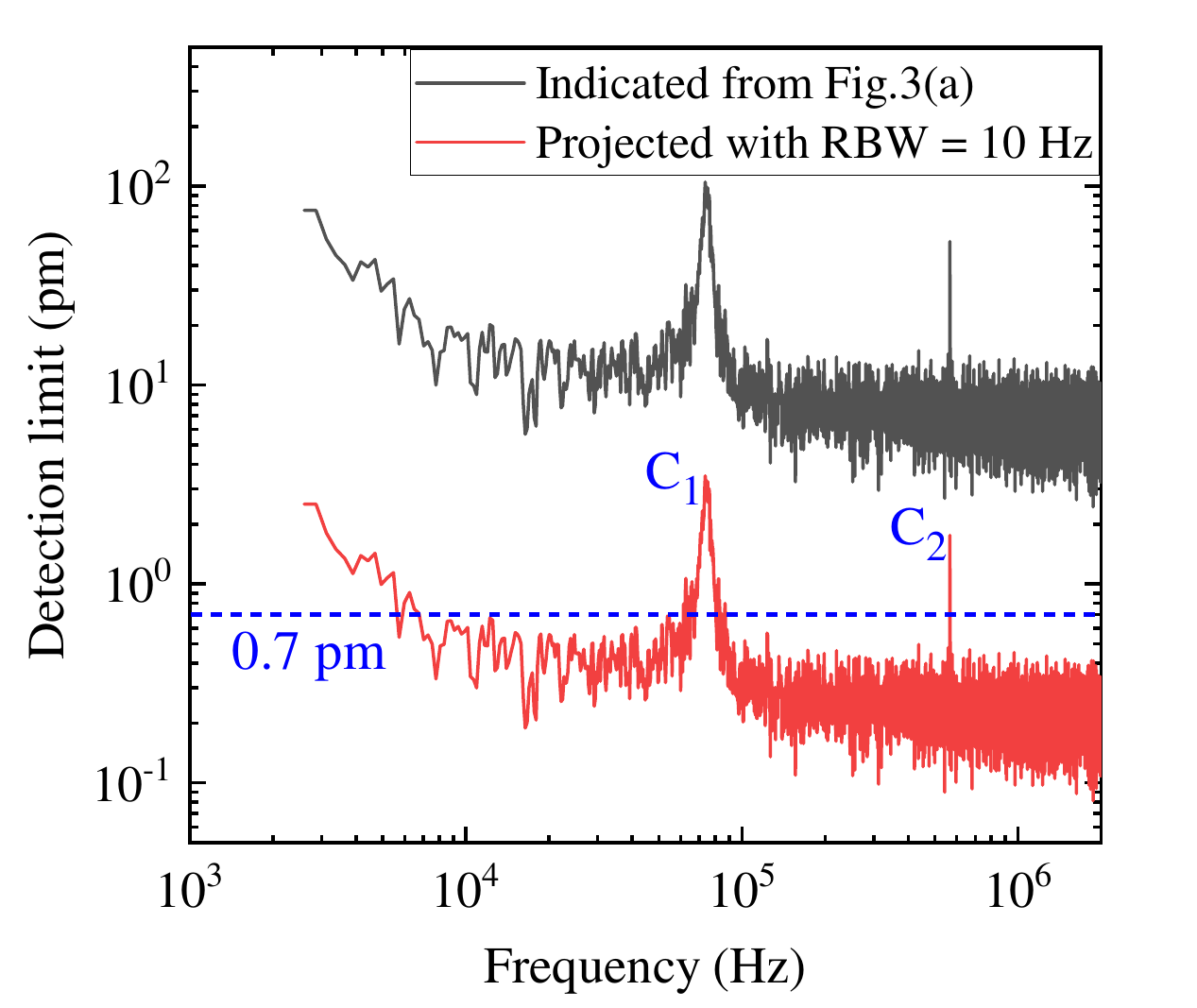}
\caption{ The detection limit of the NPRO LFI, deduced with data from Fig.3(a). The peak at 74 kHz (C\textsubscript{1}) is the sideband of RO and the peak at 567 kHz (C\textsubscript{2}) is the second harmonic of beat signal. Black curve: Indicated from Fig.3(a) with RBW of 300 Hz. Red curve: Projected with RBW of 10 Hz. The blue dashed line marks a detection level of 0.7 pm.}
\label{Fig6}
\end{figure}

To analyze the detection limit of vibration amplitude, we calculated the amplitudes with Eq.\eqref{eq15} using data from Fig.\ref{Fig3}(a), where the data trace from the right side of the beat signal is used. The result is shown as the black curve in Fig.\ref{Fig6}. Because the RBW in the measurement is 300 Hz, and the beat signal linewidth is no more than 10 Hz (as shown by the inset of Fig.\ref{Fig3}(a)), the detection limit is thus projected from the black curve by dividing 30 to lower the noise, which is shown as the red curve in Fig.\ref{Fig6}. One can see that the minimum detection limit has reached a value below 0.7 pm. It indicates that the NPRO LFI can reach the sub-picometer detection limit. Note that the two bumps of C\textsubscript{1} (74 kHz) and C\textsubscript{2} (567 kHz) are from RO noise and the second harmonic of beat signal, respectively. The detection limit is rising significantly at low-frequencies (\textless10 kHz), which is due to the instability of the beat signal induced by magnetic noise, temperature noise and pump noise, etc. A method to stabilize the beat frequency is under development. 

We analyze the accuracy of our vibration detection theoretically due to the lack of calibration tool. We first use the method shown in \citep{26} to calculate the vibration amplitude under the first-order approximate of the Bessel function, and find the inaccuracy is below 1\% when the vibration amplitude is below 24 nm. We then build a three-dimensional finite-element model with piezo and mirror to simulate the vibration amplitude under experimental parameters. We find the calculated vibration amplitude is on the same order-of-magnitude as the measurements. Thus, the measurement accuracy is indirectly verified. Note that the accuracy can also be affected by the optical path length instability induced by environment such as temperature, vibration and airflow. However, these factors are mostly influencing the detection in the low frequency range (below 1 kHz), and thus are not considered in this experiment.

\section{Conclusions}

In summary, we present a new type of laser feedback interferometry based on an NPRO laser system. Under weak magnetic intensity, the re-injected light drives an NPRO works in stable bidirectional lasing regime which constructs naturally a frequency-shifted LFI without using acoustic-optic modulators. A theoretical model based on two-frequency rate equations and Lang–Kobayashi equation is presented to show the theoretical validity of LFI in NPRO. The interference frequency-shifted signal is experimentally studied and discussed. Moreover, vibration detection is demonstrated to prove the potential application, where the detection limit can reach to sub-picometer, and the detection frequency range is from kilohertz to a few hundred kilohertz. The vibration detection frequency range, detection limit and amplitude accuracy are further discussed. Benefiting from its tiny footprint, rigid structure and excellent detection resolution, the NPRO-based frequency-shifted LFI system may find important applications in vibration, velocity, displacement and rotation measurements with high sensitivities.

\section*{Appendix A}
The simulation combines two models together in order to provide the whole spectrum of NPRO laser feedback system. The feedback model we propose in the main text contains two terms (Eqs.\eqref{eq11}-\eqref{eq12}). The first term is considered to be constant for both steady solutions $I_{s1,2}$, and the second term carries the interference signal with frequency shift. We perform Fourier transform for Eqs.\eqref{eq11}-\eqref{eq12} and only the second term remains in the result. The parameters we used in the simulation are shown in Table A.1 and the results are shown in red solid curves in Fig.\ref{Fig3}(a) and Fig.\ref{Fig4}(a). However, the RO noise peak exists in the steady solutions  $I_{s1,2}$ of NPRO in the experiment results. To provide the noise background of the simulation in the spectrum, we add the RO model \citep{30} in frequency spectrum of the simulation, shown in red dashed curves in Fig.\ref{Fig3}(a) and Fig.\ref{Fig4}(a). The RO in frequency domain can be described as:

\setcounter{equation}{0}
\renewcommand{\theequation}{A.\arabic{equation}} 
\begin{equation}
\label{EqA.1}
\begin{aligned}
V_{RO}(\omega) = 
& \frac{4\kappa_m \kappa \gamma_t^2 p (p-1)}{\left(\omega_r^2 - \omega^2 + \frac{\omega^2 \gamma \gamma_l}{\gamma^2 + \omega^2}\right)^2 + \frac{\omega^2 \gamma_l^2 \gamma^4}{(\gamma^2 + \omega^2)^2}} \\
& + \left(1 + \frac{4\kappa_m^2 (\omega^2 + \gamma_t^2) + 8\kappa \kappa_m \gamma_t \gamma_l (p-1)}{\left(\omega_r^2 - \omega^2 + \frac{\omega^2 \gamma \gamma_l}{\gamma^2 + \omega^2}\right)^2 + \frac{\omega^2 \gamma_l^2 \gamma^4}{(\gamma^2 + \omega^2)^2}}\right) \\
& + V_{P} \times \frac{4\kappa_m \kappa \gamma_t^2 (p-1)}{\left(\omega_r^2 - \omega^2 + \frac{\omega^2 \gamma \gamma_l}{\gamma^2 + \omega^2}\right)^2 + \frac{\omega^2 \gamma_l^2 \gamma^4}{(\gamma^2 + \omega^2)^2}} \\
& + \frac{4\kappa_m \kappa (\omega^2 + \gamma_l^2)}{\left(\omega_r^2 - \omega^2 + \frac{\omega^2 \gamma \gamma_l}{\gamma^2 + \omega^2}\right)^2 + \frac{\omega^2 \gamma_l^2 \gamma^4}{(\gamma^2 + \omega^2)^2}} \\
& + \frac{4\kappa_m \kappa_l (\omega^2 + \gamma_l^2)}{\left(\omega_r^2 - \omega^2 + \frac{\omega^2 \gamma \gamma_l}{\gamma^2 + \omega^2}\right)^2 + \frac{\omega^2 \gamma_l^2 \gamma^4}{(\gamma^2 + \omega^2)^2}}
\end{aligned}
\end{equation}
Here, $\omega_r=\sqrt{2\kappa \gamma_t (p-1)}$ is the frequency of RO, $V_{P}$ is the amplitude of the absorbed pumping, $p = \frac{P_{\text{in}}}{P_{\text{th}}}$ is pumping rate with $P_{\text{in}}$ representing pumping power and $P_{\text{th}}$ representing threshold power of NPRO. $\gamma_t$ and $\gamma$ are the decay rate of atomic spontaneous emission from different energy levels and $\gamma_{l} \approx p \gamma_{t}$. $\kappa_m$ and $\kappa_l$ represent laser cavity damping rates due to the output mirror and other losses, respectively. The total laser cavity damping rate is $\kappa = \kappa_m + \kappa_l$. To preserve consistency for the experiment in the noise spectrum, the relative coefficient $V = \frac{4n\kappa_m\omega_r^2}{\kappa G}$ is introduced in the simulation, where $n$ is the atomic number of interaction medium and $G$ is proportional to the stimulated emission cross-section. 

Noticed that there are also steady terms  $I_{s1,2}$ shown in the second terms of Eqs.\eqref{eq11}-\eqref{eq12}, causing modulation sidebands of RO in interference signal. Thus, the final formula in our simulation of the RO noise background is:

\renewcommand{\theequation}{A.\arabic{equation}}
\begin{equation}
V_{\text{simulation}} = \frac{V_{RO}(\omega) + A \cdot V_{RO}(\omega-\Delta\omega)}{V}
\label{eq:V_simulation}
\end{equation}
where $\Delta\omega$ is the frequency difference between the two modes as shown in Eqs.\eqref{eq11}-\eqref{eq12}, and $A$ is the fitting coefficient for the sidebands.

In the final spectrum, we do the addition of two models according to Eqs.\eqref{eq11}-\eqref{eq12} in frequency domain, and remove the region of the feedback model that is overwhelmed by the RO noise background. Two models are distinguished with different types of curves in the simulation results.

\renewcommand{\thetable}{A.\arabic{table}}
\begin{table}[htbp]
\centering
\caption{Parameters in the simulation}
\begin{tabular}{lrr} 
\toprule
Parameters for feedback model\\ 
\midrule
$\gamma_{B}$ & 5.7 × 10$^{9}$ & s$^{-1}$ \\
$\gamma_{ext}$ & 6.6 × 10$^{8}$ & s$^{-1}$ \\
$\Delta\omega$ & 5.67 × 10$^{5}$ & Hz \\
$\omega _{vib}$ & 2$\pi$ × 100 & Hz\\
$A_{vib}$ & 7 × 10$^{-9}$ &m \\
$L$ & 0.5 & m \\
$r$ & 2.3 & \\
\midrule
Parameters for RO model    \\
\midrule
$\gamma_{t}$      &  2.5 × 10$^{3}$ & s$^{-1}$ \\
$\gamma_{l}$     & 3.16 × 10$^{3}$ & s$^{-1}$ \\
$\gamma$        & 6 × 10$^{4}$ & s$^{-1}$ \\
$\kappa_m$     & 5 × 10$^{5}$ & s$^{-1}$ \\ 
$\kappa_l$     & 4.2 × 10$^{6}$ & s$^{-1}$ \\ 
$P_{\text{in}}$ &  152 & mW \\
$P_{\text{th}}$ &  120 & mW \\
$\omega_r$  & 74.43 & kHz \\
$ V_p $      &  4× 10$^{4}$ & \\
$n$      & 10$^{10}$& \\
$G$       & 10$^{20}$& \\
$A$         & 0.04 & \\
\midrule
Parameters for Fourier transform  &  \\
\midrule
Data length & 4 × 10$^{7}$ & \\
Sample rate & 10$^{10}$ & Hz\\
Frequency resolution & 300 & Hz\\
\bottomrule
\end{tabular}
\end{table}

\section*{Funding}
This work was supported by National Natural Science Foundation of China (12404489), Fundamental Research Funds for the Central Universities, Sun Yat-sen University (24QNPY162), National Key Research and Development Program of China (2020YFC2200200) and Major Projects of Basic and Applied Basic Research in Guangdong Province (2019B030302001).

\section*{CRediT authorship contribution statement}
\textbf{Rong Zhu}: Writing-original Draft, Writing-review \& editing, Investigation, Methodology, Formal analysis, Data curation. \textbf{Xuezhen Gong}: Writing-original Draft, Writing-review \& editing, Investigation, Methodology, Formal analysis, Data curation. \textbf{Wenxun Li}: Formal analysis, Writing-review \& editing. \textbf{Guobin Zhou}: Formal analysis. \textbf{Weitong Fan}: Formal analysis. \textbf{Danqing Liu}: Investigation. \textbf{Chunzhao Ma}: Investigation. \textbf{Jie Xu}: Investigation, Formal analysis, Writing-review \& editing. \textbf{Changlei Guo}: Writing-review \& editing, Formal analysis, Methodology, Supervision, Resources, Project administration, Funding acquisition. \textbf{Hsien-Chi Yeh}: Supervision, Resources, Investigation, Funding acquisition.

\section*{Declaration of competing interest}
The authors declare that they have no known competing financial interests or personal relationships that could have appeared to influence the work reported in this paper.

\section*{Data availability}
Data will be made available on request.

\bibliographystyle{elsarticle-num}

\bibliography{Reference.bib}

\begin{thebibliography}{10}

\bibitem{15}
Meiyu Chen, Fang Xie, Yuji Yang, and Xiaoyuan Zhang.
\newblock Heterodyne self-mixing interferometry to large step height measurement based on a dual-wavelength single-longitudinal-mode optical fiber laser.
\newblock {\em Opt. Laser Technol.}, 167:109821, 2023.

\bibitem{22}
S~De, V~Pal, A~El~Amili, Guillaume Pillet, G~Baili, Mehdi Alouini, I~Sagnes, R~Ghosh, and F~Bretenaker.
\newblock Intensity noise correlations in a two-frequency {VECSEL}.
\newblock {\em Opt. Express}, 21(3):2538--2550, 2013.

\bibitem{10}
S.~Donati.
\newblock Developing self‐mixing interferometry for instrumentation and measurements.
\newblock {\em Laser Photonics Rev.}, 6(3):393--417, 2012.

\bibitem{25}
Silvano Donati and Ray-Hua Horng.
\newblock The diagram of feedback regimes revisited.
\newblock {\em IEEE J. Sel. Top. Quantum Electron.}, 19(4):1500309--1500309, 2013.

\bibitem{24}
Weitong Fan, Chunzhao Ma, Danqing Liu, Rong Zhu, Guobin Zhou, Xuezhen Gong, Shungao Zhou, Jie Xu, Wenhao Yuan, Changlei Guo, and Hsien-Chi Yeh.
\newblock Dual-frequency fundamental-mode npro laser for low-noise microwave generation.
\newblock {\em Opt. Express}, 31(8):13402--13413, 2023.

\bibitem{4}
Yu~Han, Ke~Kou, Cuo Wang, and Zewei Song.
\newblock A wavelength analyzer using laser self-mixing interferometry.
\newblock {\em Opt. Lasers Eng.}, 183:108505, 2024.

\bibitem{11}
Olivier Jacquin, Eric Lacot, Wilfried Glastre, Olivier Hugon, and Hugues~Guillet de~Chatellus.
\newblock Experimental comparison of autodyne and heterodyne laser interferometry using an {N}d:{YVO}\textsubscript{4} microchip laser.
\newblock {\em J. Opt. Soc. Am. A}, 28(8):1741--1746, Aug 2011.

\bibitem{18}
T~J. Kane and R~L. Byer.
\newblock Monolithic, unidirectional single-mode {N}d:{YAG} ring laser.
\newblock {\em Opt. Lett.}, 10(2):65--67, 1985.

\bibitem{14}
E.~Lacot, R.~Day, and F.~Stoeckel.
\newblock Coherent laser detection by frequency-shifted optical feedback.
\newblock {\em Phys. Rev. A}, 64(4):469--469, 2001.

\bibitem{12}
Eric Lacot, Olivier Jacquin, Gr\'{e}goire Roussely, Olivier Hugon, and Hugues~Guillet de~Chatellus.
\newblock Comparative study of autodyne and heterodyne laser interferometry for imaging.
\newblock {\em J. Opt. Soc. Am. A}, 27(11):2450--2458, Nov 2010.

\bibitem{23}
R.~Lang and K.~Kobayashi.
\newblock External optical feedback effects on semiconductor injection laser properties.
\newblock {\em IEEE J. Quantum Electron.}, 16(3):347--355, 1980.

\bibitem{16}
Jiyang Li, Haisha Niu, and Yan~Xiong Niu.
\newblock Laser feedback interferometry and applications: a review.
\newblock {\em Opt. Eng.}, 56(5):050901--050901, 2017.

\bibitem{26}
Hanne Martinussen, Astrid Aksnes, and Helge~E Engan.
\newblock Wide frequency range measurements of absolute phase and amplitude of vibrations in micro- and nanostructures by optical interferometry.
\newblock {\em Opt. Express}, 15(18):11370--11384, 2007.

\bibitem{19}
A.C. Nilsson, E.K. Gustafson, and R.L. Byer.
\newblock Eigenpolarization theory of monolithic nonplanar ring oscillators.
\newblock {\em IEEE J. Quantum Electron.}, 25(4):767--790, 1989.

\bibitem{20}
Alan~C Nilsson, Thomas~J Kane, and Robert~L Byer.
\newblock {Monolithic Nonplanar Ring Lasers: Resistance To Optical Feedback}.
\newblock {\em Proc. SPIE}, 912:13--19, 1988.

\bibitem{27}
Lucas Oliverio, Damien Rontani, and Marc Sciamanna.
\newblock Nonlinear dynamics of a laser diode subjected to both optical injection and optical feedback.
\newblock {\em Opt. Express}, 32(15):25906--25918, Jul 2024.

\bibitem{13}
K.~Otsuka.
\newblock Effects of external perturbations on {L}i{N}d{P}\textsubscript{4}{O}\textsubscript{12} lasers.
\newblock {\em IEEE J. Quantum Electron.}, 15(7):655--663, 1979.

\bibitem{28}
Xiaoqiong Qi, Gary Agnew, Thomas Taimre, She Han, Yah~Leng Lim, Karl Bertling, Aleksandar Demi\'{c}, Paul Dean, Dragan Indjin, and Aleksandar~D. Raki\'{c}.
\newblock Laser feedback interferometry in multi-mode terahertz quantum cascade lasers.
\newblock {\em Opt. Express}, 28(10):14246--14262, May 2020.

\bibitem{17}
Song Qiu, Tong Liu, Zhengliang Liu, You Ding, Ruoyu Tang, Xiangyang Zhu, Ke~Wang, and Yuan Ren.
\newblock Self-mixing rotational doppler effect for spinning velocity detection.
\newblock {\em Opt. Laser Technol.}, 174:110721, 2024.

\bibitem{30}
Timothy~C. Ralph, Charles~C. Harb, and Hans-A. Bachor.
\newblock Intensity noise of injection-locked lasers: Quantum theory using a linearized input-output method.
\newblock {\em Phys. Rev. A}, 54:4359--4369, Nov 1996.

\bibitem{29}
Carlo Silvestri, Xiaoqiong Qi, Thomas Taimre, and Aleksandar~D. Raki\ifmmode~\acute{c}\else \'{c}\fi{}.
\newblock Multimode dynamics of terahertz quantum cascade lasers: Spontaneous and actively induced generation of dense and harmonic coherent regimes.
\newblock {\em Phys. Rev. A}, 106:053526, Nov 2022.

\bibitem{2}
Thomas Taimre, Milan Nikoli{\'c}, Karl Bertling, Yah~Leng Lim, Thierry Bosch, and Aleksandar~D Raki{\'c}.
\newblock Laser feedback interferometry: a tutorial on the self-mixing effect for coherent sensing.
\newblock {\em Adv. Opt. Photonics}, 7(3):570--631, 2015.

\bibitem{3}
Yuanfu Tan, Mubasher Ali, Feng Lin, Zhou Su, Wei-Hsin Liao, and Hay Wong.
\newblock Study on laser spot size measurement by scanning-slit method based on back-injection interferometry.
\newblock {\em Opt. Laser Technol.}, 172:110472, 2024.

\bibitem{Tian:24}
Mingwang Tian, Xin Xu, Sihong Chen, Zhipeng Feng, and Yidong Tan.
\newblock Ultrasensitive detection of remote acoustic vibrations at 300 m distance by optical feedback enhancement.
\newblock {\em Photon. Res.}, 12(9):1962--1971, Sep 2024.

\bibitem{5}
Zhanwu Xie, Jie Li, Dongmei Guo, Wei Xia, Haitao Yan, and Ming Wang.
\newblock All-fiber laser self-mixing interferometry for signal enhancement with phase-shifted fiber bragg grating.
\newblock {\em Opt. Laser Technol.}, 172:110496, 2024.

\bibitem{9}
Zhong Xu, Jiyang Li, Shulian Zhang, Yidong Tan, and Songlin Zhuang.
\newblock Remote eavesdropping at 200 meters distance based on laser feedback interferometry with single-photon sensitivity.
\newblock {\em Opt. Lasers Eng.}, 141(1):106562, 2021.

\bibitem{7}
Yuanyang Zhao, Desheng Zhu, Youze Chen, Yourui Tu, and Liang Lu.
\newblock All-fiber self-mixing laser doppler velocimetry with much less than 0.1pw optical feedback based on adjustable gain.
\newblock {\em Opt. Lett.}, 45(13):3565--3568, 2020.

\bibitem{6}
Borui Zhou, Yu~Wang, Bing Zhou, Xueju Shen, and Yidong Tan.
\newblock Highly sensitive interferometry with strong anti laser jamming capability based on frequency shift optical feedback.
\newblock {\em Opt. Laser Technol.}, 171:110449, 2024.

\bibitem{21}
Guobin Zhou, Rong Zhu, Chunzhao Ma, Xuezhen Gong, Weitong Fan, Shungao Zhou, Jie Xu, Changlei Guo, and Hsien-Chi Yeh.
\newblock Unidirectional operation criterion in monolithic nonplanar ring oscillators.
\newblock {\em Opt. Lett.}, 48(11):3047--3050, 2023.

\bibitem{8}
D.~Zhu, Y.~Zhao, Y.~Tu, H.~Li, L.~Xu, B.~Yu, and L.~Lu.
\newblock All-fiber laser feedback interferometer using a {DBR} fiber laser for effective sub-picometer displacement measurement.
\newblock {\em Opt. Lett.}, 46(1):114--117, 2021.

\end{thebibliography}

\end{document}